\documentclass[aps,prl,twocolumn,superscriptaddress,showpacs,preprintnumbers,amsmath,amssymb]{revtex4}
\usepackage{graphicx} % Include figure files
\usepackage{dcolumn}  % Align table columns on decimal point
\usepackage{rotating}
\usepackage{threeparttable}
\usepackage[colorlinks,linkcolor=red,anchorcolor=green,citecolor=blue]{hyperref}

\graphicspath{{ps}}

\newcommand{\BR}{{\cal B}}
\newcommand{\EE}{e^+e^-}
\newcommand{\pp}{\pi^+\pi^-}
\newcommand {\tabincell}[2]{\begin{tabular}{@{}#1@{}}#2\end{tabular}}%
\begin{document}

%\vspace*{-3\baselineskip}
%\resizebox{!}{3cm}{\includegraphics{belle.eps}}

%\input{pub505}

%\preprint{\vbox{ \hbox{   }
                        %\hbox{Belle DRAFT {\it 17-07}}
%                        \hbox{Intended for {\it PRL}}
%                        \hbox{Author: Y. B. Li, C. P. Shen, C. Z. Yuan}
%                        \hbox{Committee: J. Yelton(chair),}
%                        \hbox{Y. Kato, U. Tamponi }
%                        \hbox{Belle Preprint \# 2017-24}
%                        \hbox{KEK Preprint \# 2017-35}
%}}

\title{ %\quad\\[1.0cm]
Observation of $\Xi_{c}(2930)^0$ and updated measurement of $B^{-} \to K^{-} \Lambda_{c}^{+} \bar{\Lambda}_{c}^{-}$ at Belle
}

%%%% >>>>> insert the authorlist here. BEFORE the abstract !!!!! <<<<<
%%%% >>>>> from the authorship confirmation web page
%%% Name the file author.tex and use \input{author} to insert into your latex file.
%\author{Author}\affiliation{affiliation}
%\collaboration{The Belle Collaboration}
%\noaffiliation
%% end author list

%%% Paper:    X_c(2930) etc
%%% Journal:  Physical Review Letters
%%% Contacts: Y.B. Li (liyb@pku.edu.cn)
%%%           C.P. Shen (shencp@buaa.edu.cn)
%%%           C.Z. Yuan (yuancz@mail.ihep.ac.cn)
%%% Non-responding authors or those who said NO are commented out.
%%% ====================================================================
%%% Click the RELOAD button on your web browser to see the updated file.
%%% ====================================================================
%%% Use \input{author} to insert this material into your latex file.
%%%%% Force institutions to appear in alphabetical order when typeset.
\noaffiliation
%%%\affiliation{Aligarh Muslim University, Aligarh 202002}
\affiliation{University of the Basque Country UPV/EHU, 48080 Bilbao}
\affiliation{Beihang University, Beijing 100191}
%%%\affiliation{University of Bonn, 53115 Bonn}
\affiliation{Budker Institute of Nuclear Physics SB RAS, Novosibirsk 630090}
\affiliation{Faculty of Mathematics and Physics, Charles University, 121 16 Prague}
%%%\affiliation{Chiba University, Chiba 263-8522}
\affiliation{Chonnam National University, Kwangju 660-701}
\affiliation{University of Cincinnati, Cincinnati, Ohio 45221}
\affiliation{Deutsches Elektronen--Synchrotron, 22607 Hamburg}
\affiliation{University of Florida, Gainesville, Florida 32611}
%%%\affiliation{Department of Physics, Fu Jen Catholic University, Taipei 24205}
\affiliation{Fudan University, Shanghai 200443}
%%%\affiliation{Justus-Liebig-Universit\"at Gie\ss{}en, 35392 Gie\ss{}en}
\affiliation{Gifu University, Gifu 501-1193}
%%%\affiliation{II. Physikalisches Institut, Georg-August-Universit\"at G\"ottingen, 37073 G\"ottingen}
\affiliation{SOKENDAI (The Graduate University for Advanced Studies), Hayama 240-0193}
\affiliation{Gyeongsang National University, Chinju 660-701}
\affiliation{Hanyang University, Seoul 133-791}
\affiliation{University of Hawaii, Honolulu, Hawaii 96822}
\affiliation{High Energy Accelerator Research Organization (KEK), Tsukuba 305-0801}
\affiliation{J-PARC Branch, KEK Theory Center, High Energy Accelerator Research Organization (KEK), Tsukuba 305-0801}
%%%\affiliation{Hiroshima Institute of Technology, Hiroshima 731-5193}
\affiliation{IKERBASQUE, Basque Foundation for Science, 48013 Bilbao}
%%%\affiliation{University of Illinois at Urbana-Champaign, Urbana, Illinois 61801}
\affiliation{Indian Institute of Science Education and Research Mohali, SAS Nagar, 140306}
\affiliation{Indian Institute of Technology Bhubaneswar, Satya Nagar 751007}
\affiliation{Indian Institute of Technology Guwahati, Assam 781039}
\affiliation{Indian Institute of Technology Hyderabad, Telangana 502285}
\affiliation{Indian Institute of Technology Madras, Chennai 600036}
\affiliation{Indiana University, Bloomington, Indiana 47408}
\affiliation{Institute of High Energy Physics, Chinese Academy of Sciences, Beijing 100049}
\affiliation{Institute of High Energy Physics, Vienna 1050}
\affiliation{Institute for High Energy Physics, Protvino 142281}
%%%\affiliation{Institute of Mathematical Sciences, Chennai 600113}
\affiliation{University of Mississippi, University, Mississippi 38677}
\affiliation{INFN - Sezione di Napoli, 80126 Napoli}
\affiliation{INFN - Sezione di Torino, 10125 Torino}
\affiliation{Advanced Science Research Center, Japan Atomic Energy Agency, Naka 319-1195}
\affiliation{J. Stefan Institute, 1000 Ljubljana}
\affiliation{Kanagawa University, Yokohama 221-8686}
\affiliation{Institut f\"ur Experimentelle Kernphysik, Karlsruher Institut f\"ur Technologie, 76131 Karlsruhe}
%%%\affiliation{Kavli Institute for the Physics and Mathematics of the Universe (WPI), University of Tokyo, Kashiwa 277-8583}
\affiliation{Kennesaw State University, Kennesaw, Georgia 30144}
\affiliation{King Abdulaziz City for Science and Technology, Riyadh 11442}
\affiliation{Department of Physics, Faculty of Science, King Abdulaziz University, Jeddah 21589}
\affiliation{Korea Institute of Science and Technology Information, Daejeon 305-806}
\affiliation{Korea University, Seoul 136-713}
\affiliation{Kyoto University, Kyoto 606-8502}
\affiliation{Kyungpook National University, Daegu 702-701}
\affiliation{\'Ecole Polytechnique F\'ed\'erale de Lausanne (EPFL), Lausanne 1015}
\affiliation{P.N. Lebedev Physical Institute of the Russian Academy of Sciences, Moscow 119991}
\affiliation{Faculty of Mathematics and Physics, University of Ljubljana, 1000 Ljubljana}
\affiliation{Ludwig Maximilians University, 80539 Munich}
\affiliation{Luther College, Decorah, Iowa 52101}
\affiliation{University of Malaya, 50603 Kuala Lumpur}
\affiliation{University of Maribor, 2000 Maribor}
\affiliation{Max-Planck-Institut f\"ur Physik, 80805 M\"unchen}
\affiliation{School of Physics, University of Melbourne, Victoria 3010}
%%%\affiliation{Middle East Technical University, 06531 Ankara}
\affiliation{University of Miyazaki, Miyazaki 889-2192}
\affiliation{Moscow Physical Engineering Institute, Moscow 115409}
\affiliation{Moscow Institute of Physics and Technology, Moscow Region 141700}
\affiliation{Graduate School of Science, Nagoya University, Nagoya 464-8602}
\affiliation{Kobayashi-Maskawa Institute, Nagoya University, Nagoya 464-8602}
%%%\affiliation{Nara University of Education, Nara 630-8528}
\affiliation{Nara Women's University, Nara 630-8506}
\affiliation{National Central University, Chung-li 32054}
\affiliation{National United University, Miao Li 36003}
\affiliation{Department of Physics, National Taiwan University, Taipei 10617}
\affiliation{H. Niewodniczanski Institute of Nuclear Physics, Krakow 31-342}
%%%\affiliation{Nippon Dental University, Niigata 951-8580}
\affiliation{Niigata University, Niigata 950-2181}
%%%\affiliation{University of Nova Gorica, 5000 Nova Gorica}
\affiliation{Novosibirsk State University, Novosibirsk 630090}
\affiliation{Osaka City University, Osaka 558-8585}
%%%\affiliation{Osaka University, Osaka 565-0871}
\affiliation{Pacific Northwest National Laboratory, Richland, Washington 99352}
\affiliation{Panjab University, Chandigarh 160014}
\affiliation{Peking University, Beijing 100871}
%%%\affiliation{University of Pittsburgh, Pittsburgh, Pennsylvania 15260}
%%%\affiliation{Punjab Agricultural University, Ludhiana 141004}
%%%\affiliation{Research Center for Electron Photon Science, Tohoku University, Sendai 980-8578}
%%%\affiliation{Research Center for Nuclear Physics, Osaka University, Osaka 567-0047}
\affiliation{Theoretical Research Division, Nishina Center, RIKEN, Saitama 351-0198}
%%%\affiliation{RIKEN BNL Research Center, Upton, New York 11973}
%%%\affiliation{Saga University, Saga 840-8502}
\affiliation{University of Science and Technology of China, Hefei 230026}
%%%\affiliation{Seoul National University, Seoul 151-742}
%%%\affiliation{Shinshu University, Nagano 390-8621}
\affiliation{Showa Pharmaceutical University, Tokyo 194-8543}
\affiliation{Soongsil University, Seoul 156-743}
%%%\affiliation{University of South Carolina, Columbia, South Carolina 29208}
\affiliation{Stefan Meyer Institute for Subatomic Physics, Vienna 1090}
\affiliation{Sungkyunkwan University, Suwon 440-746}
\affiliation{School of Physics, University of Sydney, New South Wales 2006}
\affiliation{Department of Physics, Faculty of Science, University of Tabuk, Tabuk 71451}
\affiliation{Tata Institute of Fundamental Research, Mumbai 400005}
%%%\affiliation{Excellence Cluster Universe, Technische Universit\"at M\"unchen, 85748 Garching}
\affiliation{Department of Physics, Technische Universit\"at M\"unchen, 85748 Garching}
\affiliation{Toho University, Funabashi 274-8510}
%%%\affiliation{Tohoku Gakuin University, Tagajo 985-8537}
\affiliation{Department of Physics, Tohoku University, Sendai 980-8578}
\affiliation{Earthquake Research Institute, University of Tokyo, Tokyo 113-0032}
\affiliation{Department of Physics, University of Tokyo, Tokyo 113-0033}
\affiliation{Tokyo Institute of Technology, Tokyo 152-8550}
\affiliation{Tokyo Metropolitan University, Tokyo 192-0397}
%%%\affiliation{Tokyo University of Agriculture and Technology, Tokyo 184-8588}
\affiliation{University of Torino, 10124 Torino}
%%%\affiliation{Utkal University, Bhubaneswar 751004}
\affiliation{Virginia Polytechnic Institute and State University, Blacksburg, Virginia 24061}
\affiliation{Wayne State University, Detroit, Michigan 48202}
%%%\affiliation{Yamagata University, Yamagata 990-8560}
\affiliation{Yonsei University, Seoul 120-749}
  \author{Y.~B.~Li}\affiliation{Peking University, Beijing 100871}\affiliation{Beihang University, Beijing 100191} % Peking
   \author{C.~P.~Shen}\affiliation{Beihang University, Beijing 100191} % Beihang
% \author{A.~Abdesselam}\affiliation{Department of Physics, Faculty of Science, University of Tabuk, Tabuk 71451} % Tabuk
  \author{I.~Adachi}\affiliation{High Energy Accelerator Research Organization (KEK), Tsukuba 305-0801}\affiliation{SOKENDAI (The Graduate University for Advanced Studies), Hayama 240-0193} % KEK
% \author{K.~Adamczyk}\affiliation{H. Niewodniczanski Institute of Nuclear Physics, Krakow 31-342} % Krakow
  \author{J.~K.~Ahn}\affiliation{Korea University, Seoul 136-713} % Korea
  \author{H.~Aihara}\affiliation{Department of Physics, University of Tokyo, Tokyo 113-0033} % Tokyo
  \author{S.~Al~Said}\affiliation{Department of Physics, Faculty of Science, University of Tabuk, Tabuk 71451}\affiliation{Department of Physics, Faculty of Science, King Abdulaziz University, Jeddah 21589} % Tabuk
% \author{K.~Arinstein}\affiliation{Budker Institute of Nuclear Physics SB RAS, Novosibirsk 630090}\affiliation{Novosibirsk State University, Novosibirsk 630090} % BINP
% \author{Y.~Arita}\affiliation{Graduate School of Science, Nagoya University, Nagoya 464-8602} % Nagoya
  \author{D.~M.~Asner}\affiliation{Pacific Northwest National Laboratory, Richland, Washington 99352} % PNNL
% \author{H.~Atmacan}\affiliation{University of South Carolina, Columbia, South Carolina 29208} % SouthCarolina
% \author{V.~Aulchenko}\affiliation{Budker Institute of Nuclear Physics SB RAS, Novosibirsk 630090}\affiliation{Novosibirsk State University, Novosibirsk 630090} % BINP
  \author{T.~Aushev}\affiliation{Moscow Institute of Physics and Technology, Moscow Region 141700} % MIPT
  \author{R.~Ayad}\affiliation{Department of Physics, Faculty of Science, University of Tabuk, Tabuk 71451} % Tabuk
% \author{T.~Aziz}\affiliation{Tata Institute of Fundamental Research, Mumbai 400005} % Tata
  \author{V.~Babu}\affiliation{Tata Institute of Fundamental Research, Mumbai 400005} % Tata
% ...
% BINP
% \author{M.~Ziegler}\affiliation{Institut f\"ur Experimentelle Kernphysik, Karlsruher Institut f\"ur Technologie, 76131 Karlsruhe} % Karlsruhe
% \author{T.~Zivko}\affiliation{J. Stefan Institute, 1000 Ljubljana} % Ljubljana
% \author{A.~Zupanc}\affiliation{Faculty of Mathematics and Physics, University of Ljubljana, 1000 Ljubljana}\affiliation{J. Stefan Institute, 1000 Ljubljana} % Ljubljana
% \author{N.~Zwahlen}\affiliation{\'Ecole Polytechnique F\'ed\'erale de Lausanne (EPFL), Lausanne 1015} % Lausanne
\collaboration{The Belle Collaboration}

\begin{abstract}

We report the first observation of the $\Xi_{c}(2930)^0$ charmed-strange baryon with a significance
greater than 5$\sigma$. The $\Xi_{c}(2930)^0$ is found in its decay to $K^- \Lambda_{c}^+$ in $B^{-} \to K^{-} \Lambda_{c}^{+}
\bar{\Lambda}_{c}^{-}$ decays. The measured mass and width are
$[2928.9 \pm 3.0(\rm stat.)^{+0.9}_{-12.0}(\rm syst.)]$~MeV/$c^{2}$ and $[19.5 \pm 8.4(\rm stat.)
^{+5.9}_{-7.9}(\rm syst.)]$~MeV, respectively, and the product branching fraction is
$\BR(B^{-} \to \Xi_{c}(2930)^0 \bar{\Lambda}_{c}^{-})
\BR(\Xi_{c}(2930)^0 \to K^- \Lambda_{c}^{+})=[1.73 \pm 0.45(\rm stat.) \pm 0.21(\rm syst.)]\times 10^{-4}$. We also measure $\BR(B^{-} \to K^{-} \Lambda_{c}^{+}
\bar{\Lambda}_{c}^{-}) = [4.80 \pm 0.43(\rm stat.) \pm 0.60(\rm syst.)] \times 10^{-4}$ with
improved precision,
and search for the charmonium-like state $Y(4660)$ and its spin
partner, $Y_{\eta}$, in the $\Lambda_{c}^{+}\bar{\Lambda}_{c}^{-}$
invariant mass spectrum.
No clear signals of the $Y(4660)$ nor its spin partner
are observed and the 90\% credibility level (C.L.) upper limits on their production rates are determined.
These measurements are obtained from a sample of
$(772\pm11)\times 10^{6} B\bar{B}$ pairs collected at the
$\Upsilon(4S)$ resonance by the Belle detector at the KEKB
asymmetric energy electron-positron collider.
\end{abstract}

 %13.30.-a,13.25.-k,}

\maketitle

%%%% >>>> keep the final version single-spaced
\tighten

{\renewcommand{\thefootnote}{\fnsymbol{footnote}}}
\setcounter{footnote}{0}
%\linenumbers
%%%%%%%%%%%%%%%%%%%%%%%%%%%%%%%%%%%%%%
%%%%%%%%%   introduction  %%%%%%%%%%%%
%%%%%%%%%%%%%%%%%%%%%%%%%%%%%%%%%%%%%%

The singly charmed baryon is composed of a charm quark and two
light quarks. Charmed baryon spectroscopy provides an excellent
ground for studying the dynamics of light quarks in the
environment of a heavy quark and offers an excellent laboratory
for testing heavy-quark or chiral symmetry of the heavy or light quarks, respectively.
Although many new excited charmed
baryons have been discovered by BaBar, Belle, CLEO and LHCb in the
past two decades~\cite{PDG}, and many efforts have been made to identify the
quantum numbers of these new states and understand their
properties, we do not yet have
a fully phenomenological model that describes the complicated physics of
this sector~\cite{chenghy1, chenghy2}.
%Further experimental studies on excited charmed baryons are
%needed.
Identification and observation of new members in the charmed-baryon family
will provide more information to address these open issues.

The $\Xi_{c}(2930)$ charmed-strange baryon has been reported only
in the analysis of  $B^{-} \to K^{-} \Lambda_{c}^{+}
\bar{\Lambda}_{c}^{-}$ by
BaBar~\cite{barbar}, where they claim  a signal in the
 $K^-\Lambda_{c}^+$ invariant mass
distribution with a  mass of $[2931
\pm 3({\rm stat.}) \pm 5({\rm syst.})]$ MeV/$c^{2}$ and a width of
$[36 \pm 7({\rm stat.}) \pm 11 ({\rm syst.})]$ MeV. However,
neither the results of the fit to their spectrum nor the significance of the signal were given; the Particle Data
Group (PDG) lists it as a ``one star" state~\cite{PDG}.
Despite the weak experimental evidence for the $\Xi_{c}(2930)$ state, it has been taken into
account in many theoretical models, including the chiral quark model~\cite{lei}, the light-cone
Quantum Chromodynamics (QCD) sum rule~\cite{huaxing}, the $^3P_0$ mode~\cite{Xi-spectrum}, the constituent quark model~\cite{kai,bing},
and the heavy-hadron chiral perturbation theory~\cite{Hai}.

Belle has previously studied $B^{-} \to K^{-} \Lambda_{c}^{+}
\bar{\Lambda}_{c}^{-}$ decays~\cite{belleold} with a data sample of $386\times
10^{6}~B\bar{B}$ pairs
but the distributions of the intermediate $K\Lambda_c$ systems have
not been presented.
The full Belle data sample of $(772\pm11)\times
10^{6} B\bar{B}$ pairs
permits an improved study of $B^{-} \to
K^{-} \Lambda_{c}^{+} \bar{\Lambda}_{c}^{-}$ and a test for the existence of the $\Xi_{c}(2930)$.

%It is of great interest to notice that
The same $B$ decay mode can be used to study
the $\Lambda_{c}^{+} \bar{\Lambda}_{c}^{-}$
invariant mass. In this system, Belle has previously observed a charmonium-like state,
the $Y(4630)$, in the initial state radiation (ISR) process
$\EE \to \gamma_{\rm ISR} \Lambda_{c}^{+}
\bar{\Lambda}_{c}^{-}$~\cite{4630} with a measured mass of
$[4634^{+8}_{-7}({\rm stat.})^{+5}_{-8}({\rm syst.})]$ MeV/$c^2$
and a width of $[92^{+40}_{-24}({\rm stat.}) ^{+10}_{-21}({\rm
syst.})]$ MeV. As this mass is very close to that of the $Y(4660)$ observed by Belle
in the ISR process $e^+e^- \to \gamma_{\rm
ISR}\pp\psi'$~\cite{Belle4660isr-old,Belle4660isr-new}, many
theoretical explanations assume they are the same
state~\cite{2Ysame1,2Ysame2,4616b}.
In Refs.~\cite{thero4600,4616a}, where the $Y(4660)$ is modeled as an
$f_{0}(980)\psi'$ bound state,
the authors predict that it should have a spin partner---a $f_{0}(980)\eta_{c}(2S)$ bound state denoted as the
$Y_{\eta}$---with a mass and width of $(4613 \pm 4)$ MeV/$c^2$ and around 30
MeV, respectively, and a large partial width into $\Lambda_{c}^{+} \bar{\Lambda}_{c}^{-}$~\cite{4616a,4616b}.

%In its analysis of $B^{-} \to K^{-} \Lambda_{c}^{+}
%\bar{\Lambda}_{c}^{-}$, BaBar provided the $\Lambda_{c}^{+} \bar{\Lambda}_{c}^{-}$
%invariant mass distribution, and two small peaks can be seen~\cite{barbar}.
%These may correspond to the $Y_{\eta}$ and
%$Y(4660)$ but there is also a large probability that they are just
%statistical fluctuations or reflections of charmed baryon states.

In this Letter, we perform an updated measurement of $B^{-} \to
K^{-} \Lambda_{c}^{+} \bar{\Lambda}_{c}^{-}$~\cite{charge-conjugate} and
observe the $\Xi_{c}(2930)^0$ signal with a
significance of 5.1$\sigma$. This analysis is based on the full data sample
collected at the
$\Upsilon(4S)$ resonance by the Belle detector~\cite{Belle} at the
KEKB asymmetric energy electron-positron collider~\cite{KEKB}.
Simulated signal events with $B$ meson decays are generated using
{\sc EvtGen}~\cite{evtgen}, while the inclusive decays are generated
via  {\sc PYTHIA}~\cite{pythia}. These events are  processed by a
detector simulation based on
{\sc GEANT3}~\cite{geant3}. Inclusive Monte Carlo (MC) samples
of $\Upsilon(4S)\to B \bar{B}$ ($B=B^+$ or $B^0$) and $e^+e^- \to q \bar{q}$
($q=u,~d,~s,~c$) events at $\sqrt{s}=10.58$ GeV are used to
check the backgrounds, corresponding to more than 5 times the
integrated luminosity of the data.

%%%%%%%%%%%%%%%%%%%%%%%%%%%%%%%%%%%%%
%%%%%%%%   Event selection %%%%%%%%%%
%%%%%%%%%%%%%%%%%%%%%%%%%%%%%%%%%%%%%

We reconstruct the $\Lambda_c^+$ via the  $\Lambda_c^+\to p K^- \pi^+$,
$pK_{S}^{0}$, $\Lambda \pi^{+}$, $pK_{S}^{0}\pi^{+}\pi^{-}$, and
$\Lambda \pi^{+}\pi^{+}\pi^{-}$ decay channels. When a
$\Lambda_c^+$ and $\bar{\Lambda}_{c}^{-}$  are combined
to reconstruct a $B$ candidate, at least one is required to
have been reconstructed via the $p K^+ \pi^-$ or $\bar{p} K^- \pi^+$ decay
process.  For charged tracks, information from different detector
subsystems, including specific ionization in the central drift chamber, time
measurements in the time-of-flight scintillation counters and the response of the
aerogel threshold Cherenkov counters, is combined to
form the likelihood ${\mathcal L}_i$ for species $i$,
where $i=\pi$,~$K$, or $p$~\cite{pid}. Except for the charged tracks
from $\Lambda \to p \pi^{-}$ and $K_{S}^{0} \to \pi^{+} \pi^{-}$
decays, a track with a likelihood ratio $\mathcal{R}_K^{\pi} =
\mathcal{L}_K/(\mathcal{L}_K + \mathcal{L}_\pi)> 0.6$ is
identified as a kaon, while a track with $\mathcal{R}_K^{\pi}<0.4$ is
treated as a pion~\cite{pid}. With this selection, the kaon (pion)
identification efficiency is about 94\% (98\%), while 5\% (2\%) of
the kaons (pions) are misidentified as pions (kaons). A track with
$\mathcal{R}^\pi_{p/\bar{p}} =
\mathcal{L}_{p/\bar{p}}/(\mathcal{L}_{p/\bar{p}}+\mathcal{L}_\pi)
> 0.6$ and $\mathcal{R}^K_{p/\bar{p}} =
\mathcal{L}_{p/\bar{p}}/(\mathcal{L}_{p/\bar{p}}+\mathcal{L}_K) >
0.6$ is identified as a proton/anti-proton with an efficiency of
about 98\%; fewer than $1\%$ of the pions/kaons are misidentified as protons/anti-protons.

%Charged tracks reconstructed in the CDC, except for tracks from $\Lambda \to p \pi^{-}$ and $K_{s}^{0} \to \pi^{+} \pi^{-}$ decays, are required to originate from the interaction point(IP).
%We distinguish charged kaons, pions and protons based on a kaon(pion,proton) likelihood ($\mathcal{L}_{K(\pi,p)}$) derived from the time-of-flight scintillation counters, ACC, and dE/dx measurements in the CDC.
%We reconstruct $B^{-} \to K^{-} \Lambda_{c}^{+} \bar{\Lambda}_{c}^{-}$ candidates with $\Lambda_c$ decay in to $pK^{-}\pi^{+}$, $pK_{s}^{0}$, $\Lambda \pi^{+}$, $pK_{s}^{0}\pi^{+}\pi^{-}$ and $\Lambda \pi^{+}\pi^{+}\pi^{-}$. At least one of $\Lambda_{c}$ is required to have been reconstructed via the $pK\pi$ decay process and the inclusion of the charge conjugate mode decay is implied.

The $K_{S}^{0}$ candidates are reconstructed from pairs of
oppositely-charged tracks, treated as pions, and identified by a
multivariate analysis with a neural network~\cite{NN} based on two
sets of input variables~\cite{NN-input}. Candidate $\Lambda$
baryons are reconstructed in the decay $\Lambda \to p \pi^-$ and
selected if the $p \pi^-$ invariant mass is within 5 MeV/$c^2$
(5$\sigma$) of the $\Lambda$ nominal mass~\cite{PDG}.

We perform a vertex fit to signal $B$ candidates.
If there is more than one $B$ signal candidate in an event,
we select the one with the minimum $\chi^{2}_{\rm
vertex}$ from the vertex fit. We require $\chi^{2}_{\rm
vertex}<50$ with a selection efficiency above 96\%.
As the continuum background level is very low, continuum suppression is not necessary.

The $B$ candidates are identified using the beam-energy
constrained mass $M_{\rm bc}$ and the mass difference $\Delta M_{B}$.
The beam-energy constrained mass is defined as $M_{\rm bc} \equiv
\sqrt{E_{\rm beam}^{2}/c^2 - (\sum \vec{p}_{i})^2}/c$, where
$E_{\rm beam}$ is the beam energy and $\vec{p}_{i}$ are
the three-momenta of the $B$-meson decay products, all defined in
the center-of-mass system (CMS) of the $\EE$ collision. The mass
difference is defined as $\Delta M_{B} \equiv M_{B} -
m_{B}$, where $M_{B}$ is the invariant mass of the $B$
candidate and $m_{B}$ is the nominal $B$-meson
mass~\cite{PDG}.

%%%%%%%%%%%%%%%%%%%%%%%%%%%%%%%%%%%%%
%%%%%%% analysis plots  %%%%%%%%%%%%%
%%%%%%%%%%%%%%%%%%%%%%%%%%%%%%%%%%%%%

\begin{figure}[htbp]
\begin{center}
\includegraphics[width=4.36cm]{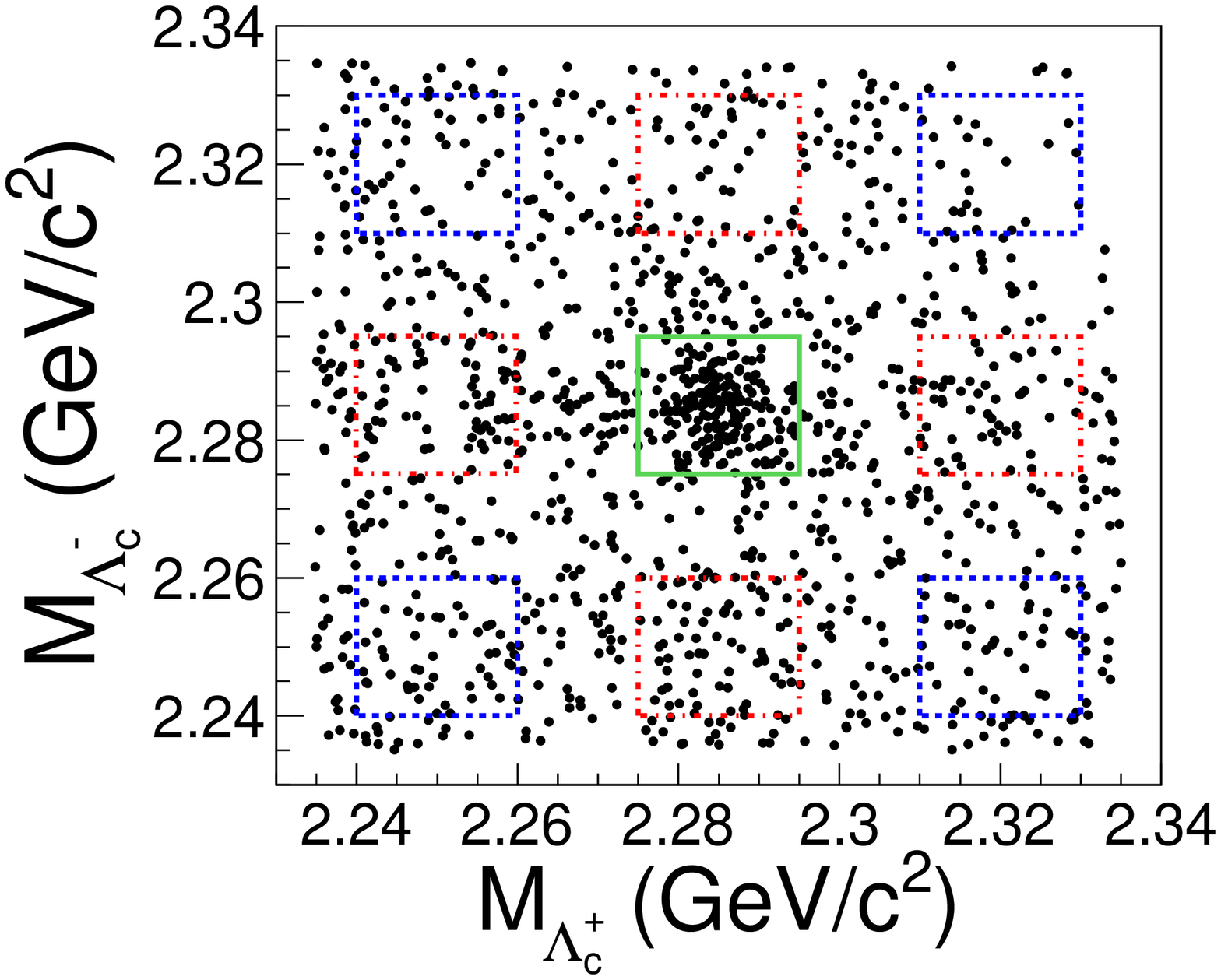}
\includegraphics[width=4.17cm]{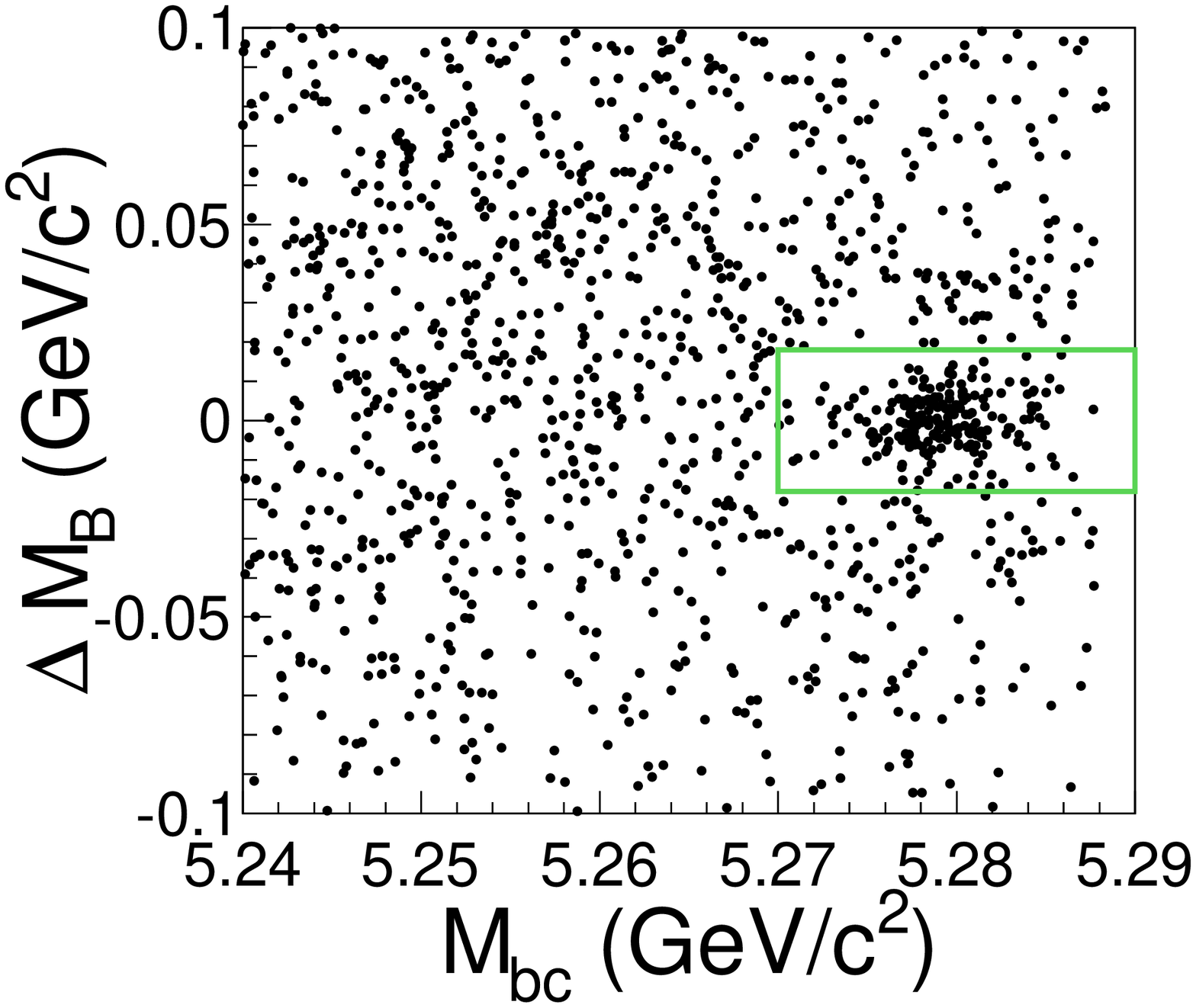}
\caption{\label{comlbdac_bsignal}
Signal-enhanced distribution  of
$M(\bar{\Lambda}_{c}^{-})$ versus $M(\Lambda_{c}^{+})$ (left
panel) and of $\Delta M_{B}$ versus $M_{\rm bc}$ (right panel)
from the selected $B^{-} \to K^{-} \Lambda_{c}^{+}
\bar{\Lambda}_{c}^{-}$ candidates,
summing over all five reconstructed $\Lambda_{c}$ decay modes.
Each panel shows the events falling in the solid green
signal region of the other panel.
The dashed red and blue boxes in the left panel show the $\Lambda_c$ sideband regions used for the estimation of the non-$\Lambda_c$ background (see text).}
\end{center}
\end{figure}

Figure~\ref{comlbdac_bsignal}
shows clear evidence of $\Lambda_c^+$ and $\bar{\Lambda}_c^-$ in the distribution
of $M_{\bar{\Lambda}_{c}^{-}}$ versus
$M_{\Lambda_{c}^{+}}$ (left panel) from the selected $B^{-} \to K^{-}
\Lambda_{c}^{+} \bar{\Lambda}_{c}^{-}$ data candidates
in the $B$ signal region
of $|\Delta M_B| < 0.018\,{\rm GeV}/c^2$ and $M_{\rm bc} > 5.27\,{\rm GeV}/c^2$ ($\sim 3\sigma$),
illustrated by the green box in the right panel's distribution of $\Delta M_B$ versus $M_{\rm bc}$.
The $\Lambda_{c}$ signal region
(the central green box in the left panel) is defined as $|M_{\Lambda_{c}} -
m_{\Lambda_{c}} |< 10$ MeV/$c^{2}$ ($ \sim 2.5\sigma$), where
$m_{\Lambda_{c}}$ is the nominal mass of the $\Lambda_{c}$
baryon~\cite{PDG}.
As the mass resolution of $\Lambda_{c}$
candidates is almost independent of the $\Lambda_{c}$ decay mode,
according to the signal MC simulation,
the same requirement is placed on all $\Lambda_{c}$ decay modes.
The non-$\Lambda_c$ background in the $\Lambda_c$ signal region
is estimated as half of the total number of events in the four red sideband regions
minus one quarter of the total number of events in the four blue sideband regions of the left panel.
%To estimate the non-$\Lambda_c$
%backgrounds, we define the $\Lambda_{c}^{+}$ and
%$\bar{\Lambda}_{c}^{-}$ mass sidebands as the half of the events
%in solid line boxes minus quarter of events in dashed line boxes,
%as shown in Fig.~\ref{comlbdac_bsignal} (left panel).

%After all event selection requirements, the scatter plot of
%$\Delta M_{B}$ versus $M_{\rm bc}$ from the selected $B^{-} \to K^{-}
%\Lambda_{c}^{+} \bar{\Lambda}_{c}^{-}$ data candidates
%summing over all five reconstructed $\Lambda_c$ decay modes is shown in
%Fig.~\ref{comlbdac_bsignal} (right panel),
%in which clear $B$ signals are observed.

  %%%%%%%%%%%%%%%%%%%%%%%%%%%%%%%%%%%
  %*********  B signal   ***********%
  %%%%%%%%%%%%%%%%%%%%%%%%%%%%%%%%%%%

To obtain the $B^{-} \to K^{-} \Lambda_{c}^{+}
\bar{\Lambda}^{-}_{c}$ signal yields, we perform an unbinned
two-dimensional (2D) simultaneous extended maximum likelihood fit
to the $\Delta M_{B}$ versus $M_{\rm bc}$ distributions for the
five reconstructed $\Lambda_{c}$ decay modes. The model used to fit the $M_{\rm bc}$ distribution is a
Gaussian function for the signal shape plus an ARGUS
function~\cite{argus} for the background. The model for the $\Delta M_{B}$
distribution is the sum of a Gaussian function for the signal plus a first-order
polynomial for the background. The Gaussian
resolutions  are fixed to the
values from the fits to the individual MC distributions, and the relative signal yields
among the five final states is fixed according to
the relative branching fraction
between the final states and the detection acceptance and efficiency of the intermediate states.

\begin{figure}[htbp]
\begin{center}
\includegraphics[width=4.2cm]{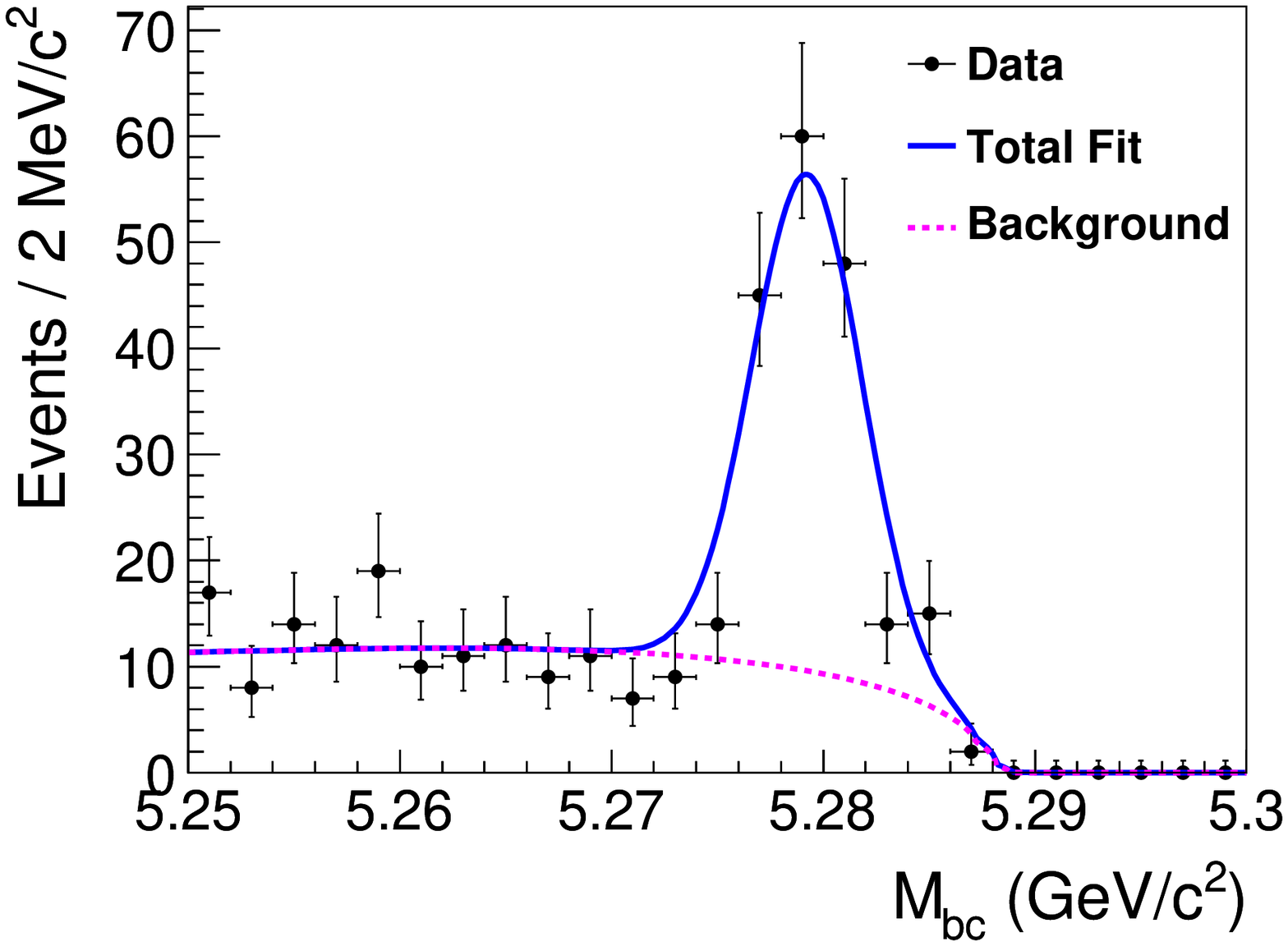}
\includegraphics[width=4.2cm]{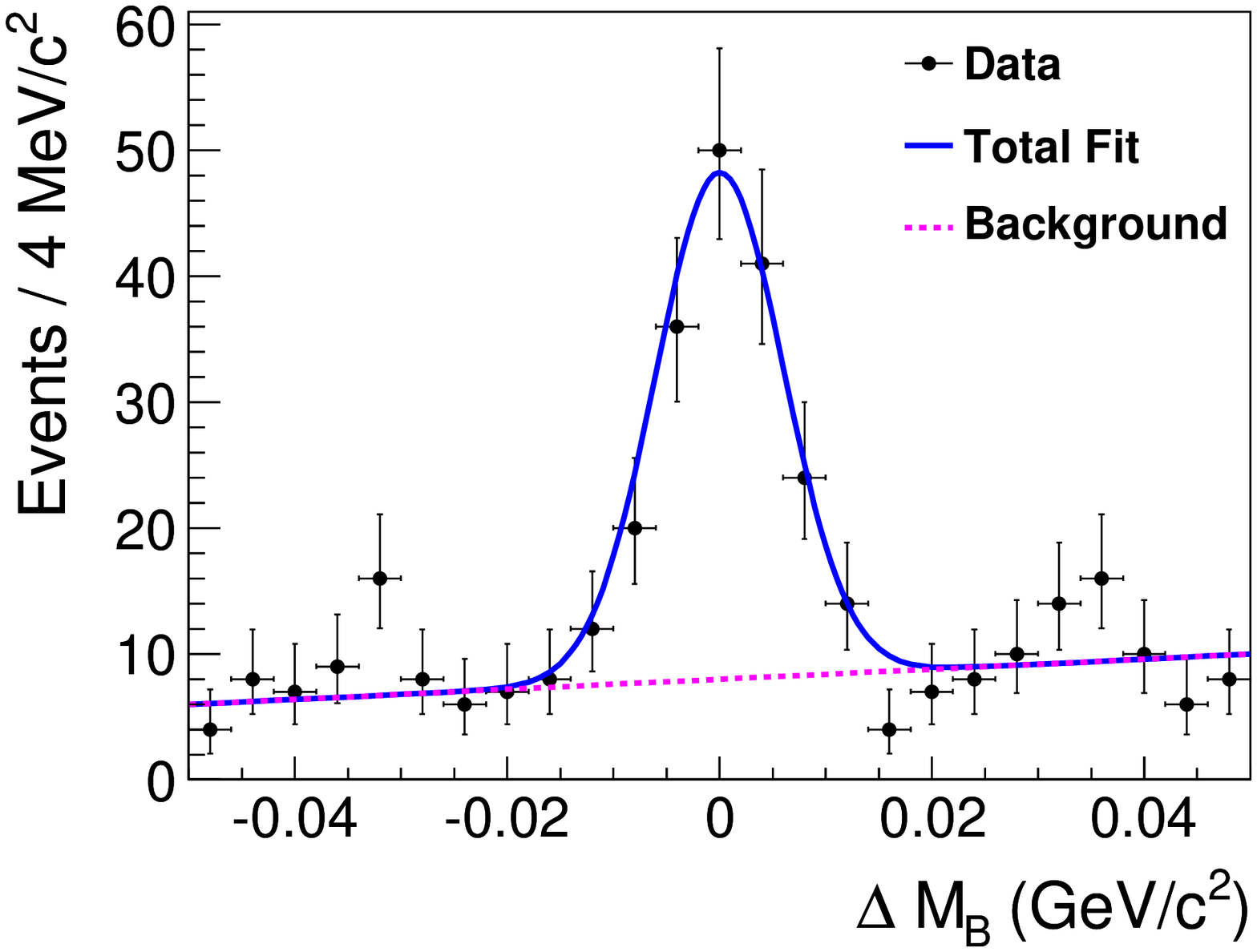}
\put(-210,70){\bf (a)}
 \put(-90,70){\bf (b)}
\caption{\label{MvsEcom} The $\Lambda_c$-signal-enhanced distributions of (a) $M_{\rm bc}$ in the $\Delta
M_{B}$ signal region and (b) $\Delta M_{B}$ in the $M_{\rm bc}$ signal
region  for $B^{-} \to K^{-} \Lambda_{c}^{+}
\bar{\Lambda}^{-}_{c}$, combining five exclusive final states. The
dots with error bars are data, the solid blue curves are the best-fit
projections to the distributions and the dashed magenta  lines are the fitted
backgrounds.}
\end{center}
\end{figure}

Figure~\ref{MvsEcom} shows the projections of the fit
superimposed on the $\Lambda_c$-signal-enhanced  $M_{\rm bc}$ and $\Delta M_{B}$ distributions,
summing over all five reconstructed $\Lambda_{c}$ decay modes.
We observe  $153 \pm 14$ signal events with a signal significance above 10$\sigma$, and extract the branching
fraction of $\BR (B^{-} \to K^{-}
\Lambda_{c}^{+} \bar{\Lambda}_{c}^{-})=[4.80 \pm 0.43({\rm
stat.})] \times 10^{-4}$.

  %%%%%%%%%%%%%%%%%%%%%%%%%%%%%%%%%%%
  %***********  Xi_c ***************%
  %%%%%%%%%%%%%%%%%%%%%%%%%%%%%%%%%%%

\begin{figure}[htbp]
\begin{center}
\includegraphics[width=6cm]{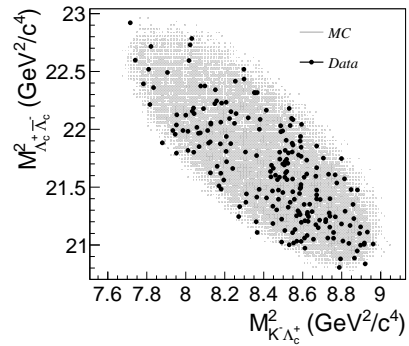}
\caption{\label{sigScat} Dalitz distribution of reconstructed $B^{-} \to
K^{-} \Lambda_{c}^{+} \bar{\Lambda}_{c}^{-}$ candidates in the $B$
signal region. The black dots are data; the shaded region is
the MC simulated phase-space distribution.}
\end{center}
\end{figure}

%We take $|\Delta M_{B}| < 0.018$ GeV/$c^{2}$ and $M_{\rm bc}
%> 5.27$ GeV/$c^{2}$ ($\sim3\sigma$) as the $B$ signal region, as indicated by
%the solid box in Fig.~\ref{comlbdac_bsignal} (right panel).
The Dalitz distribution of the reconstructed $B^{-} \to K^{-} \Lambda_{c}^{+}
\bar{\Lambda}_{c}^{-}$ candidates is shown in Fig.~\ref{sigScat}.
A vertical-band enhancement near
$M(K^-\Lambda_{c}^+)\sim 2.93$ GeV/$c^2$ is observed; no
signal band is apparent in the $M(\Lambda_c^+ \bar{\Lambda}_c^-)$ horizontal direction
nor in the $M(K^- \bar{\Lambda}_c^-)$ diagonal direction.

\begin{figure}[htbp]
\begin{center}
\includegraphics[width=6.4cm]{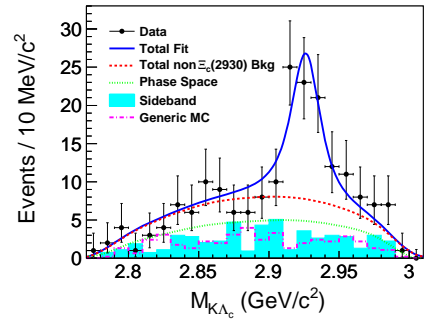}
\vspace{-0.3cm}
\caption{ \label{lkll}
The $M_{K^-\Lambda_{c}^+}$ distribution of the selected data candidates, with fit results superimposed.
Dots with error bars
are the data, the solid blue line is the best fit, the dashed red
line is the total non-$\Xi_{c}(2930)$ backgrounds, the dotted green line is the phase
space contribution, the shaded cyan  histogram is from the normalized
$\Lambda_{c}^{+}$ and $\bar{\Lambda}_{c}^{-}$ mass sidebands, and
the dot-dashed magenta  line is the sum of the MC-simulated  contributions from the
normalized $\EE \to q\bar{q}$ and $\Upsilon(4S) \to B \bar{B}$
generic-decay backgrounds.}
\end{center}
\end{figure}

The $B$-signal-enhanced  $K^{-} \Lambda_{c}^{+}$  mass spectrum is shown in Fig.~\ref{lkll}. The shaded histogram is from the
normalized $\Lambda_{c}^{+}$ and $\bar{\Lambda}_{c}^{-}$ mass
sidebands, and the dot-dashed line is the sum of the contributions
from normalized $\EE \to q\bar{q}$ and $\Upsilon(4S) \to B
\bar{B}$ generic MC samples.
Since they are consistent, we take the $\Lambda_{c}^{+}$ and $\bar{\Lambda}_{c}^{-}$
mass sidebands to represent the total background, neglecting
the small possible contribution of background with real $\Lambda_{c}^{+}$ and $\bar{\Lambda}_{c}^{-}$.
A clear $\Xi_c(2930)$ signal is
observed. No structure is seen in the $\Lambda_{c}^{+}$ and $\bar{\Lambda}_{c}^{-}$ mass
sidebands, nor in the generic MC samples, nor in the wrong-sign-combination distribution of
$K^- \bar{\Lambda}_{c}^{-}$.

An unbinned simultaneous extended maximum likelihood fit is
performed to the $K^-\Lambda_{c}^+$ invariant mass spectra for
selected $B$- and $\Lambda_c$-signal events and the $\Lambda_{c}^{+}$ and
$\bar{\Lambda}_{c}^{-}$ mass sidebands. An S-wave Breit-Wigner
(BW) function convolved with a Gaussian function with the phase
space factor and efficiency curve included (the mass resolution of Gaussian function being
fixed to 4.46 MeV/$c^2$ from the signal MC simulation) is taken as the
$\Xi_c(2930)$ signal shape.
Direct three-body $B$ decays are modeled by the shape corresponding to
$B^{-} \to K^{-} \Lambda_{c}^{+} \bar{\Lambda}_{c}^{-}$
MC-simulated decays distributed uniformly in phase space.
A second-order polynomial is used to represent the
$\Lambda_{c}^{+}$ and $\bar{\Lambda}_{c}^{-}$ mass-sideband
distribution, which is normalized to represent the
total background in the signal events in the fit.

The fit results are shown in Fig.~\ref{lkll}. The fitted mass
and width of the $\Xi_{c}(2930)$ are $M_{\Xi_{c}(2930)} = [2928.9 \pm
3.0({\rm stat.})]$ MeV/$c^{2}$ and $\Gamma_{\Xi_{c}(2930)} = [19.5
\pm 8.4(\rm stat.)]$ MeV,
where a fit bias of 1.4 MeV/$c^2$ on the $\Xi_{c}(2930)$ mass, determined using MC simulation,
has been corrected.
The yields of the $\Xi_{c}(2930)$ signal
and the phase-space contribution are $N_{\Xi_{c}} = 61 \pm 16 $
and $N_{\rm phsp} = 79 \pm 19$.

%The statistical significance of
%the $\Xi_{c}(2930)$ signal is $5.1 \sigma$, calculated from the
%difference of the logarithmic likelihoods~\cite{significance},
%$-2\ln(\mathcal{L}_{0}/\mathcal{L}_{\rm max}) = 32.8$, where
%$\mathcal{L}_{0}$ and $\mathcal{L}_{\rm max}$ are the
%maximized likelihoods without and with a signal component, respectively, taking into account
%the difference in the number of degrees of freedom ($\Delta$ndf = 3).
%The signal significance remains at 5.1$\sigma$ when convolving the  likelihood with a Gaussian function whose
%width equals the total systematic uncertainty.
%Alternative fits to the $K^-\Lambda_{c}^+$ mass spectra are performed:
%(a) using a first-order or third-order polynomial as background
%shape; (b) changing the $\Xi_{c}(2930)$ mass resolution by 10\%;
%and (c) using an energy-dependent BW function as the $\Xi_{c}(2930)$
%signal shape. The $\Xi_{c}(2930)$ signal significance is
%larger than $5.0\sigma$ in all cases.

To estimate the $\Xi_{c}(2930)$ signal significance,
we use an ensemble of simulated experiments to estimate the
probability that background fluctuations alone would produce signals
as significant as that seen in the data. We generate $K^-\Lambda_{c}^+$
mass spectra according to the shape of the non-$\Xi_{c}(2930)$
background distribution (the dashed red line in Fig.~\ref{lkll}),
with each spectrum containing 192 events which corresponds to the total data entries in
Fig.~\ref{lkll}. We fit each spectrum as we do the real data,
searching for the most significant fluctuation, and thus obtain the distribution of
$-2\ln(L_0/L_{\rm max})$ for these simulated background samples.
We perform a total of 13.2 million simulations and found 3 trials
with a $-2\ln(L_0/L_{\rm max})$ value greater than or equal to the
value obtained in the data. The resulting $p$ value is $2.27\times
10^{-7}$, corresponding to a significance of $5.1\sigma$.

The product branching fraction of $\BR(B^{-} \to \Xi_{c}(2930)
\bar{\Lambda}_{c}^{-})$ $\BR(\Xi_{c}(2930) \to K^{-}
\Lambda_{c}^{+})=[1.73 \pm 0.45(\rm stat.)]\times 10^{-4}$ is
calculated as $N_{\rm total}^{\Xi_{c}}/[2N_{B^{\pm}}\varepsilon_{\rm all}^{\Xi_{c}}
\BR(\Lambda_{c}^{+} \to p K^{-} \pi^{+})^{2}]$, where
$N_{\rm total}^{\Xi_{c}}$ is the fitted $\Xi_{c}(2930)$ signal yield;
$N_{B^\pm}=N_{\Upsilon(4S)}\BR(\Upsilon(4S)\to B^+B^-)$
($N_{\Upsilon(4S)}$ is the number of accumulated $\Upsilon(4S)$ events
and $\BR(\Upsilon(4S)\to B^+B^-)=0.514\pm0.006$~\cite{PDG}); $\BR(\Lambda_c^+\to p K^-
\pi^+)=(6.35\pm0.33)\%$ is the world-average branching fraction for
$\Lambda_c^+\to p K^- \pi^+$~\cite{PDG};
$\varepsilon_{\rm all}^{\Xi_{c}} = \sum \varepsilon_{i}^{\Xi_{c}}
\times \Gamma_{i}/\Gamma(p K^{-} \pi^{+})$
($i$ is the $\Lambda_c$ decay-mode index, $\varepsilon_{i}^{\Xi_{c}}$ is
the detection efficiency from MC simulation and
$\Gamma_{i}$ is the partial decay width of $\Lambda_{c}^{+} \to p
K^{-} \pi^{+},~pK_{S}^0,~\Lambda \pi^{-},~pK_{S}^0 \pi^{+} \pi^{-}$,
and $\Lambda \pi^{-} \pi^{+} \pi^{-}$~\cite{PDG}). Here,
$\BR(K_{S}^0 \to \pp) $ or $\BR(\Lambda \to p \pi^{-})$ is included
in $\Gamma_{i}$ for the final states with a $K_{S}^0$ or a
$\Lambda$.

\begin{figure}[h]
\begin{center}
\includegraphics[width=4.2cm]{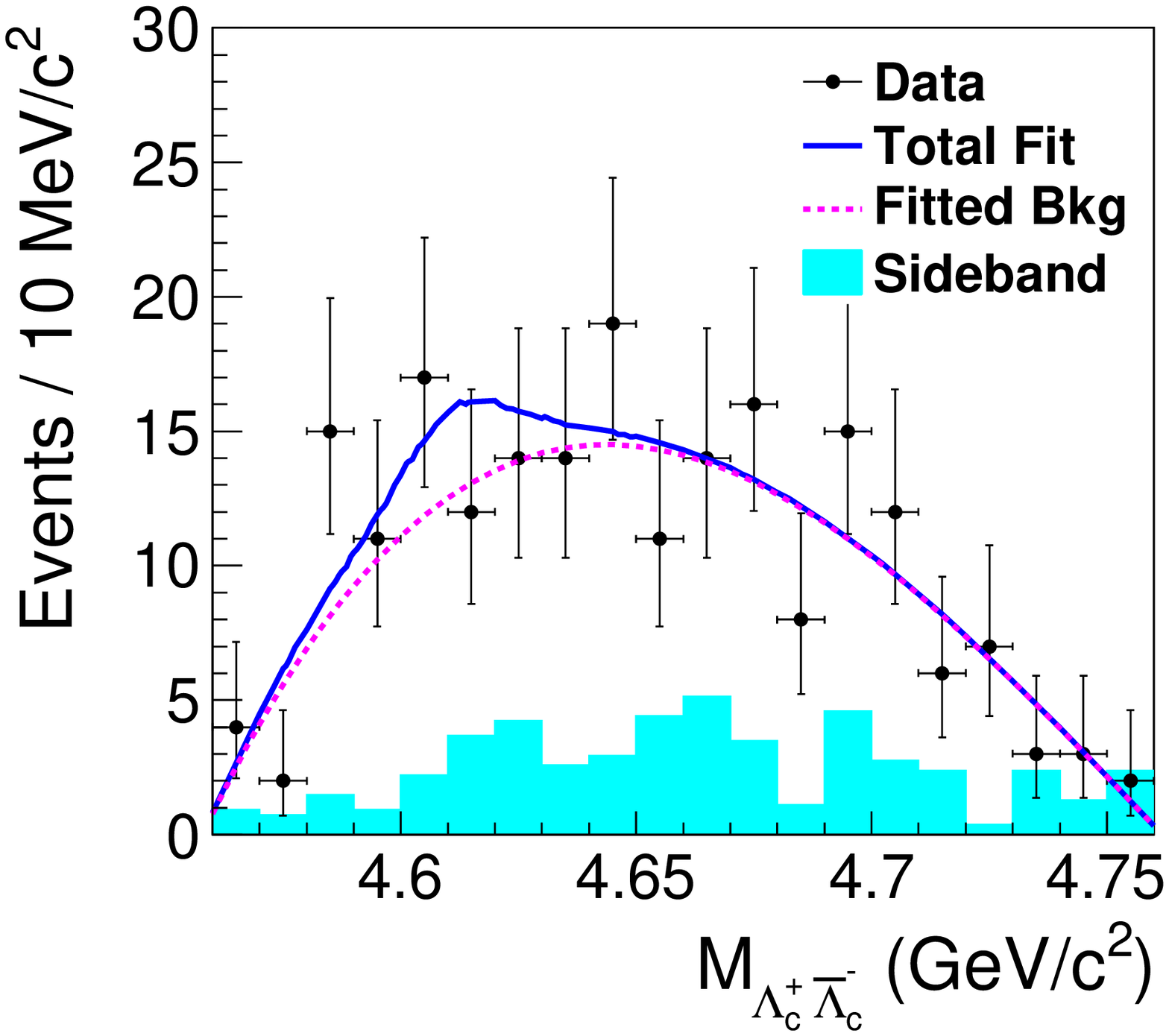}
\includegraphics[width=4.2cm]{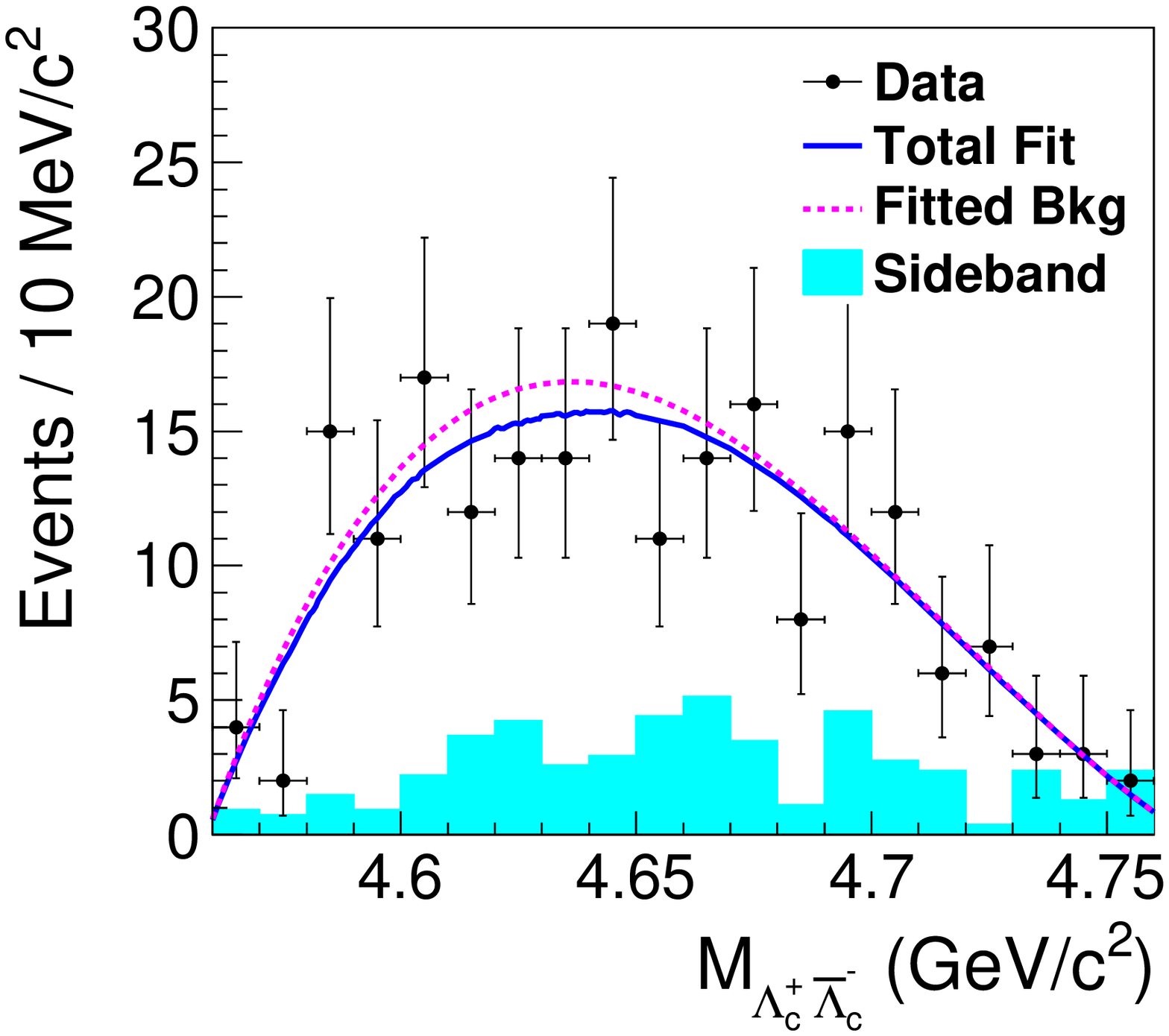}
\put(-250,80){\bf (a)}
 \put(-90,80){\bf (b)}
\caption{\label{Yplot} The $\Lambda_{c}^{+}\bar{\Lambda}_{c}^{-}$
invariant mass spectra in data with (a) $Y_{\eta}$ and (b) Y(4660)
signals included in the fits. The solid blue lines are the best fits
and the dotted red lines represent the backgrounds.
The shaded cyan  histograms are from the normalized
$\Lambda_{c}^{+}$ and $\bar{\Lambda}_{c}^{-}$ mass sidebands.
}
\end{center}
\end{figure}

The $M_{\Lambda_{c}^{+}\bar{\Lambda}_{c}^{-}}$ spectrum is shown
in Fig.~\ref{Yplot}, in which no clear $Y_{\eta}$ or $Y(4660)$ signals
is evident. An unbinned extended  maximum likelihood fit is
applied to the $\Lambda_{c}^{+}\bar{\Lambda}_{c}^{-}$ mass
spectrum to extract the signal yields of the $Y_{\eta}$ and $Y(4660)$
in $B$ decays. In the fit, the signal shape of the $Y_{\eta}$ or
$Y(4660)$ is obtained from MC simulation directly with the input
parameters $M_{Y_{\eta}} = 4616$ MeV/$c^{2}$ and
$\Gamma_{Y_{\eta}} = 30$ MeV for $Y_{\eta}$~\cite{4616b}, and
$M_{Y(4660)} = 4643$ MeV/$c^{2}$ and $\Gamma_{Y(4660)} = 72$ MeV
for $Y(4660)$~\cite{PDG};  a third-order polynomial
is used to describe all other contributions. The fit results are shown
in Figs.~\ref{Yplot}(a) and (b) for the $Y_{\eta}$ and $Y(4660)$,
respectively. From the fits, we have
$(10\pm 23)$ $Y_{\eta}$ signal events with a statistical signal
significance of $0.7\sigma$, and $(-10 \pm  26)$ $Y(4660)$ signal
events.

As the statistical signal significance of each $Y$ state is less
than $3\sigma$, 90\% C.L. Bayesian upper limits on $\BR(B^{-} \to K^{-}
Y)\BR( Y \to \Lambda_{c}^{+} \bar{\Lambda}_{c}^{-})$ are determined  to be
$1.2 \times 10^{-4}$ and $2.0 \times 10^{-4}$ for $Y=Y_\eta$ and
$Y(4660)$, respectively, by solving the equation
$\int_{0}^{\BR^{\rm
up}}\mathcal{L}(\BR)d\BR/\int_0^{+\infty}\mathcal{L}(\BR) d\BR =
0.9$, where $\BR = n_{Y}/[2\varepsilon_{\rm all}^{Y}
N_{B^{\pm}}\BR(\Lambda_{c}^{+} \to p K^{-} \pi^{+})^{2}]$ is the
assumed product branching fraction; $\mathcal{L}(\BR)$ is the
corresponding maximized likelihood of the data; $n_{Y}$ is the
number of $Y$ signal events; and $\varepsilon_{\rm all}^{Y} = \sum
\varepsilon_{i}^{Y} \times \Gamma_{i}/\Gamma(p K^{-}
\pi^{+})$ ($\varepsilon_{i}^{Y}$ being the total efficiency from MC simulation for mode $i$).
To take the systematic
uncertainty into account, the above likelihood is convolved with a
Gaussian function whose width equals the total systematic
uncertainty.

%%%%%%%%%%%%%%%%%%%%%%%%%%%%%%%%%%%%%
%%%%%%  systematic error  %%%%%%%%%%%
%%%%%%%%%%%%%%%%%%%%%%%%%%%%%%%%%%%%%

%There are several sources of systematic errors for the branching fraction measurements and the $\Xi_{c}(2930)$ parameters. Tracking efficiency uncertainties are estimated 0.35\% per track for charged track except for the tracks from $K_{S}^{0}$ and $\Lambda$ decays~\cite{trkerr}. For each mode, the final tracking uncertainties are summed up linearly. The uncertainty due to particle identification efficiency is 1.34\% for each pion and 1.6\% for each lepton, respectively.
There are several sources of systematic uncertainties in the
branching fraction measurements. The detection efficiency relevant
(DER) errors include those for tracking efficiency (0.35\%/track), particle
identification efficiency (1.9\%/kaon, 0.9\%/pion, 2.4\%/proton and
2.0\%/anti-proton),
as well as $\Lambda$ (3.0\%) and $K_{S}^{0}$  (1.7\%) selection efficiencies. Assuming all the above
systematic error sources are independent, the DER errors are
summed in quadrature for each decay mode, yielding 5.8--8.3\%,
depending on the mode. For the four branching
fraction measurements, the final DER errors are summed in
quadrature over the five $\Lambda_c$ decay modes
using weight factors equal to the product of the total efficiency and the $\Lambda_c$ partial decay width.
We estimate the systematic errors associated with
the fitting procedure by changing the order of the background
polynomial, the range of the fit, and the values of the masses and
widths of the $Y_{\eta}$ and $Y(4660)$ by $\pm 1\sigma$, and by
enlarging the mass resolution by 10\%; the
deviations from nominal in the
fitted results are taken as systematic errors.
Uncertainties for $\BR(\Lambda_c^+ \to p K^- \pi^+)$ and $\Gamma_{i}/\Gamma(p
K^{-} \pi^{+})$ are taken from Ref.~\cite{PDG}. The
final errors on the $\Lambda_c$ partial decay widths are
summed in quadrature over the five modes with the
detection efficiency as a weighting factor.
The world average of $\BR(\Upsilon(4S)\to B^+B^-)$ is $(51.4\pm0.6)\%$~\cite{PDG},
which corresponds to a systematic uncertainty of 1.2\%.
The systematic uncertainty on $N_{\Upsilon(4S)}$
is 1.37\%.
Assuming all sources listed in Table~\ref{tab:err2} to be independent, the total systematic uncertainties on the
branching fraction measurements are summed in quadrature.

\begin{table}[htbp]
	\begin{threeparttable}
		\caption{\label{tab:err2}  Relative systematic
			uncertainties (\%) in the branching fraction measurements. Here,
			$\BR_1\equiv \BR (B^{-} \to K^{-} \Lambda_{c}^{+} \bar{\Lambda}_{c}^{-})$, $\BR_2 \equiv \BR(B^{-}
			\to \Xi_{c}(2930) \bar{\Lambda}_{c}^{-})\BR(\Xi_{c}(2930) \to K^- \Lambda_{c}^+)$, $\BR_3 \equiv
			\BR(B^- \to K^- Y_{\eta})\BR( Y_{\eta} \to \Lambda_{c}^{+} \bar{\Lambda}_{c}^{-} )$, and $\BR_4 \equiv
			\BR(B^- \to K^- Y(4660))\BR( Y(4660) \to \Lambda_{c}^{+} \bar{\Lambda}_{c}^{-} )$.
		}
		\begin{tabular}{c||c|c|c|c|cp{3cm}}
			\hline\hline
			Branching fraction & DER &  Fit  &  \tabincell{c}{$\Lambda_{c}$\\ decays} & \tabincell{c}{ $N_{B^{\pm}}$}  & \tabincell{c}{Sum} \\
			\hline
			$\BR_1$         & 4.81  & 3.94 & 10.81 & 1.82 & 12.6 \\
			$\BR_2$         & 4.73  & 2.27 & 10.81 & 1.82 & 12.1 \\
			$\BR_3$         & 4.76  & 8.65 & 10.86 & 1.82 & 14.8 \\
			$\BR_4$         & 4.77  & 23.1 & 10.83 & 1.82 & 26.0 \\
			\hline\hline
		\end{tabular}
		%    \end{minipage}
	\end{threeparttable}
\end{table}

    %%%%%%%%%%%%%%%%%%%%%%%%%%%%%%%%%%%%%
    %%%%%%  Xi_{c} parameters   %%%%%%%%%
    %%%%%%%%%%%%%%%%%%%%%%%%%%%%%%%%%%%%%

The following systematic uncertainties are considered for the $\Xi_{c}(2930)$ mass and width.
Half of the correction due to the fitting bias on the $\Xi_{c}(2930)$ mass is taken conservatively as a systematic error.
By enlarging the
mass resolution by 10\%, the difference in the measured
$\Xi_{c}(2930)$ width is 0.7 MeV, which is taken as a systematic
error. By changing the
background shape, the differences of 0.3 MeV/$c^2$ and 0.9 MeV in
the measured $\Xi_{c}(2930)$  mass and width, respectively, are taken as
systematic uncertainties.

The signal-parametrization systematic uncertainty is estimated by replacing the
constant total width with a  mass-dependent width of $\Gamma_{t} =
\Gamma^{0}_{t}\times\Phi(M_{K^-\Lambda_{c}^+})/\Phi(M_{\Xi_c(2930)})$, where
$\Gamma_{t}^{0}$ is the width of the resonance,
$\Phi(M_{K^-\Lambda_c^+}) = P/M_{K^-\Lambda_c^+}$ is the phase
space factor for an S-wave two-body system ($P$ is the $K^-$
momentum in the $K^- \Lambda_c^+$ CMS) and $M_{\Xi_c(2930)}$ is
the $K^- \Lambda_c^+$ invariant mass fixed at the $\Xi_{c}(2930)$
nominal mass. The differences in the measured $\Xi_{c}(2930)$ mass
and width are 0.2 MeV/$c^2$ and 5.3 MeV, respectively, which are
taken as the systematic errors.
Adding an additional possible resonance with mass and width free
at around 2.85 GeV/$c^{2}$ into the fit to the $M(K^- \Lambda_c^+)$
spectra, the fit gives $M_{\Xi_{c}(2930)} = (2929.3 \pm 3.1)$ MeV/$c^{2}$ and
$\Gamma_{\Xi_{c}(2930)} = (21.7 \pm 9.3)$ MeV;
the differences of $+0.4$ MeV/$c^{2}$ and $+2.2$ MeV
from the mass and width found without the additional resonance, respectively,
are taken as systematic errors.
An alternative fit to the $M(K^- \Lambda_c^+)$
spectra with interference between the $\Xi_{c}(2930)$ and the phase-space
contribution included gives $M_{\Xi_{c}(2930)} = (2917.0 \pm
5.5)$ MeV/$c^{2}$ and $\Gamma_{\Xi_{c}(2930)} = (13.8 \pm 6.9)$ MeV;
the differences of $-11.9$ MeV/$c^{2}$ and $-5.7$ MeV
from the nominal mass and width, respectively,
are taken as systematic errors. Assuming
all the sources are independent, we add them in quadrature to
obtain the total systematic uncertainties on the $\Xi_{c}(2930)$
mass and width of $^{+0.9}_{-12.0}$ MeV/$c^2$ and $^{+5.9}_{-7.9}$ MeV, respectively.

%%%%%%%%%%%%%%%%%%%%%%%%%%%%%%%%%%%%%
%%%%%%%%%%% summary  %%%%%%%%%%%%%%%%
%%%%%%%%%%%%%%%%%%%%%%%%%%%%%%%%%%%%%

In summary, using $(772 \pm 11) \times 10^{6}~B\bar{B}$ pairs,
we perform an updated analysis of $B^{-} \to K^{-} \Lambda_{c}^{+}
\bar{\Lambda}_{c}^{-}$. In the $K^- \Lambda_{c}^+$ mass spectrum,
the charmed baryon state $\Xi_{c}(2930)^0$ is
clearly observed for the first time with a statistical significance greater than 5$\sigma$. The measured mass and width are
$M_{\Xi_{c}(2930)} =(2928.9 \pm 3.0 ^{+0.9}_{-12.0})$~MeV/$c^{2}$
and $\Gamma_{\Xi_{c}(2930)} = (19.5 \pm 8.4^{+5.9}_{-7.9})$ MeV.
The branching fraction is $\BR(B^{-} \to K^{-} \Lambda_{c}^{+}
\bar{\Lambda}_{c}^{-})=(4.80 \pm 0.43 \pm 0.60)\times 10^{-4} $, which is consistent with the world average
value of $(6.9 \pm 2.2) \times 10^{-4}$~\cite{PDG} but with much-improved precision.
We measure the product branching fraction $\BR(B^{-} \to
\Xi_{c}(2930) \bar{\Lambda}_{c}^{-})\BR(\Xi_{c}(2930) \to
K^- \Lambda_{c}^{+})=(1.73 \pm 0.45 \pm 0.21)
\times 10^{-4}$, where the first error is statistical and the
second systematic.
Because of the limited statistics, we do not attempt
analysis of angular correlations to determine
the spin parity of
the $\Xi_{c}(2930)^0$, however we expect that this will be possible with the much larger data sample which will be
collected with the Belle II detector.
Without this information,
we are not able to identify the quark content of this state as there
are many theoretical possibilities.
%To confirm the existence of the $\Xi_{c}(2930)^0$ indirectly,
%the analysis of $B^0 \to K^0 \Lambda_{c}^{+}
%\bar{\Lambda}_{c}^{-}$ with full of Belle data sample to search for
%charged $\Xi_{c}(2930)$ state is our following work.}
There are no significant signals seen in the $\Lambda_{c}^{+} \bar{\Lambda}_{c}^{-}$ mass spectrum.
We place 90\% C.L. upper limits for the $Y(4660)$ and its theoretically predicted spin partner $Y_{\eta}$ of
$\BR(B^- \to K^- Y(4660))\BR(
Y(4660) \to \Lambda_{c}^{+} \bar{\Lambda}_{c}^{-})<1.2\times 10^{-4}$ and $\BR(B^-
\to K^- Y_{\eta})\BR( Y_{\eta} \to \Lambda_{c}^{+}
\bar{\Lambda}_{c}^{-})<2.0 \times 10^{-4}$~\cite{BBB2}.

%%%%%%%%%%%%%%%%%%%%%%%%%%%%%%%%%%%%%
%%%%%%%   Acknowledgments   %%%%%%%%%
%%%%%%%%%%%%%%%%%%%%%%%%%%%%%%%%%%%%%
%We thank the KEKB group for excellent operation of the
%accelerator; the KEK cryogenics group for efficient solenoid
%operations; and the KEK computer group, the NII, and PNNL/EMSL for
%valuable computing and SINET5 network support. We acknowledge
%support from MEXT, JSPS and Nagoya's TLPRC (Japan); ARC
%(Australia); FWF (Austria); NSFC, CAS and CCEPP (China); MSMT
%(Czechia); CZF, DFG, EXC153, and VS (Germany); DST (India); INFN
%(Italy); MOE, MSIP, NRF, RSRI, FLRFAS project and GSDC of KISTI
%(Korea); MNiSW and NCN (Poland); MES and RFAAE (Russia); ARRS
%(Slovenia); IKERBASQUE and MINECO (Spain); SNSF (Switzerland); MOE
%and MOST (Taiwan); and DOE and NSF (USA).

%----------- Long version, for most papers -----------
We thank the KEKB group for the excellent operation of the
accelerator; the KEK cryogenics group for the efficient
operation of the solenoid; and the KEK computer group,
the National Institute of Informatics, and the
PNNL/EMSL computing group for valuable computing
and SINET5 network support.  We acknowledge support from
the Ministry of Education, Culture, Sports, Science, and
Technology (MEXT) of Japan, the Japan Society for the
Promotion of Science (JSPS), and the Tau-Lepton Physics
Research Center of Nagoya University;
the Australian Research Council;
Austrian Science Fund under Grant No.~P 26794-N20;
the National Natural Science Foundation of China under Contracts
No.~10575109, No.~10775142, No.~10875115, No.~11175187, No.~11475187,
No.~11521505, No.~11575017 and No.~11761141009;
the Chinese Academy of Science Center for Excellence in Particle Physics;
Key Research Program of Frontier Sciences, Chinese Academy of Science, Grant No.~QYZDJ-SSW-SLH011;
the Ministry of Education, Youth and Sports of the Czech
Republic under Contract No.~LTT17020;
the Carl Zeiss Foundation, the Deutsche Forschungsgemeinschaft, the
Excellence Cluster Universe, and the VolkswagenStiftung;
the Department of Science and Technology of India;
the Istituto Nazionale di Fisica Nucleare of Italy;
National Research Foundation (NRF) of Korea Grants No.~2014R1A2A2A01005286, No.~2015R1A2A2A01003280,
No.~2015H1A2A1033649, No.~2016R1D1A1B01010135, No.~2016K1A3A7A09005603, No.~2016R1D1A1B02012900; Radiation Science Research Institute, Foreign Large-size Research Facility Application Supporting project and the Global Science Experimental Data Hub Center of the Korea Institute of Science and Technology Information;
the Polish Ministry of Science and Higher Education and
the National Science Center;
the Ministry of Education and Science of the Russian Federation and
the Russian Foundation for Basic Research;
the Slovenian Research Agency;
Ikerbasque, Basque Foundation for Science and
MINECO (Juan de la Cierva), Spain;
the Swiss National Science Foundation;
the Ministry of Education and the Ministry of Science and Technology of Taiwan;
and the U.S.\ Department of Energy and the National Science Foundation.


\begin{thebibliography}{99}
\bibitem{charge-conjugate} Inclusion of charge
conjugate states is implicit unless otherwise stated.
\bibitem{BBB2} Considering the possible change of $\BR(\Lambda_{c}^{+} \to p K^{-} \pi^{+})$
in the future, we provide $\BR(B^{-} \to K^{-} \Lambda_{c}^{+}
\bar{\Lambda}_{c}^{-})\BR(\Lambda_{c}^{+} \to p K^{-} \pi^{+})^{2} = (1.94 \pm 0.13 \pm 0.14)\times 10^{-6}$
and $\BR(B^{-} \to
\Xi_{c}(2930) \bar{\Lambda}_{c}^{-})\BR(\Xi_{c}(2930)\to
K^- \Lambda_{c}^{+})\BR(\Lambda_{c}^{+} \to p K^{-} \pi^{+})^{2} = (6.97 \pm 1.81 \pm 0.43)
\times 10^{-7}$, where the first errors are statistical and the
second systematic with the uncertainty on $\BR(\Lambda_{c}^{+} \to p K^{-} \pi^{+})$ omitted.
The 90\% C.L. upper limits on $\BR(B^- \to K^- Y(4660))\BR(Y(4660) \to \Lambda_{c}^{+} \bar{\Lambda}_{c}^{-})\BR(\Lambda_{c}^{+} \to p K^{-}\pi^{+})^{2}$
and $\BR(B^-
\to K^- Y_{\eta})\BR( Y_{\eta} \to \Lambda_{c}^{+}
\bar{\Lambda}_{c}^{-}) \BR(\Lambda_{c}^{+} \to p K^{-} \pi^{+})^{2}$
are $4.8\times 10^{-7}$ and $8.0 \times 10^{-7}$, respectively.
\end{thebibliography}
\end{document}